\documentclass[aps,prl,twocolumn,showpacs]{revtex4}
\usepackage{graphicx}
\begin{document}
\title{Emittance fluctuation of mesoscopic conductors in the presence of disorders}
\author{Wei Ren$^{*}$}
\affiliation{Department of Physics and the Center of Theoretical and
Computational Physics, The University of Hong Kong, Hong Kong,
China}
\author{Fuming Xu}
\affiliation{Department of Physics and the Center of Theoretical and
Computational Physics, The University of Hong Kong, Hong Kong,
China}
\author{Jian Wang$^\dagger$}
\affiliation{Department of Physics and the Center of Theoretical and
Computational Physics, The University of Hong Kong, Hong Kong,
China}
\begin{abstract}
We report the investigation of the dynamic conductance fluctuation
of disordered mesoscopic conductors including 1D, 2D and quantum dot
systems. Our numerical results show that in the quasi-ballistic
regime the average emittance is negative indicating the expected
inductive-like behavior. However, in the diffusive and localized
regime, the average emittance is still negative. This disagrees {\it
qualitatively} with the result obtained from the random matrix
theory. Our analysis suggests that this counter-intuitive result is
due to the appearance of non-diffusive elements in the system, the
necklace states (or the precursor of necklace states in the
diffusive regime) whose existence has been confirmed experimentally
in an optical system.
\end{abstract}
 \maketitle

The universal behavior of sample-to-sample fluctuation of mesoscopic
conductors has attracted intensive theoretical and experimental
studies in last two decades\cite{lee85}. For dc conductance it is
well known that the quantum interference gives rise to a
reproducible fluctuation in the coherent mesoscopic structures. When
the sample size $L$ is smaller than the coherence length but greater
than the elastic mean free path, the system is in the diffusive
regime and the conductance fluctuation exhibits a universal behavior
with a universal conductance fluctuation (UCF) that is independent
of Fermi energy, disorder strength, and system size. The numerical
results of conductance fluctuation for various systems\cite{num} are
in good agreement with the UCF values obtained from the diagrammatic
perturbation theory\cite{lee85} and the random matrix
theory\cite{beenakker}. For ac transport, due to the long range
Coulomb interaction, the displacement current should be included in
the ac current in a self-consistent manner\cite{buttiker}. Hence it
is important to examine the role played by the displacement current
in the dynamic conductance fluctuation. For chaotic quantum dots,
the fluctuating admittance and capacitance have been studied using
random matrix theory\cite{buttiker2,buttiker3}. Careful experiments
have been carried out to measure this dynamic conductance in clean
samples\cite{regul03,gabelli}. At low frequencies, the dynamic
response of a system to the external bias is characterized by
emittance which is the imaginary part of the low frequency
admittance and is proportional to the partial density of state (DOS)
of the system\cite{buttiker}. Physically, for a conductor with large
transmission coefficient, the system responds like an inductor and
the emittance is negative while for a conductor with low
transmission coefficient the emittance is positive and the system
behaves like a capacitor. For a ballistic conductor, the emittance
$E_\mu$ is found to be $E_\mu=-(1/4) e^2 dN/dE$ while for a metallic
diffusive wire the average emittance is $E_\mu=(1/6) e^2 dN/dE$
where $dN/dE$ is the total DOS of the scattering
region\cite{buttiker4}. For a chaotic quantum dot, the emittance is
also positive from the calculation of random matrix
theory\cite{buttiker2} showing capacitive-like response. For a
disordered 2D waveguide, De Jesus et al have calculated the
emittance distribution using a continuous model\cite{degesus00}. The
emittance distribution is Gaussian at weak disorder and crossover to
a non-Gaussian distribution with a tail in the negative emittance
region at strong disorder.

For low dimensional systems, it has been predicted by
Pendry\cite{pendry} that there exists quasi-extended states known as
necklace states in strongly disordered regime. These necklace states
are due to the multi-resonance and are very rare events. Recently,
the existence of the analogous optical necklace states in disordered
1D system has been confirmed experimentally in the Anderson
localized regime\cite{bertolotti}. Since these necklace states come
from multi-resonance, the electron dwells long time with large DOS
in the scattering region giving rise to a large emittance. Thus we
expect that the necklace states must have major influence on the
average emittance for a disordered 1D system. Motivated by this
observation, in this paper we report an investigation of the
disorder effects on the emittance and its distribution using
tight-binding models for a carbon nanotube system (a quasi-one
dimensional system), a two-dimensional waveguide, and a quantum dot.
Our numerical results show that
different from the previous physical expectation and theoretical
results, the average emittance $E_\mu$ is always {\it negative} even
in the localized regime. We attribute this counter-intuitive result
as due to the non-diffusive elements in the system, the existence of
necklace states (or the precursor of necklace states in the
diffusive regime).

Single wall carbon nantoube (SWNT)\cite{bock} is a promising
candidate for the ac nanoelectronic transistor.
In this work, we consider a disordered (6,6) SWNT of 480 atoms
connected to two semi-infinite perfect (6,6) SWNT leads.
We employ the conventional nearest-neighbor
tight-binding model whose Hamiltonian for the disordered SWNT reads
$H=\sum_{i}\epsilon_{i}c^\dag_ic_i+\sum_{i,j}(t_{i,j}c^\dag_ic_j+h.c.)$,
where $c^\dag_i$ ($c_i$) is the creation (annihilation) operator for
an electron on the carbon site $i$, $\epsilon_{i}=0$ and $t_{i,j}=3
eV$ represent the on-site energy and the nearest-neighbor hopping
integral, respectively. Static Anderson-type disorder is added to
$\epsilon_{i}$ with a uniform distribution in the interval
$[-W/2,W/2]$ where $W$ denotes the disorder strength.
In this study, we have neglected the strong
electron-electron interaction in SWNT.

The conductance of the nanotube is calculated via the
Landauer-Buttiker formula, $G(0)=(2e^2/h)T$ where
$T=\rm{Tr}(\Gamma_1 G^r \Gamma_2 G^a)$ is the transmission
coefficicent, $G^r=[E-H-\Sigma^r_{1}-\Sigma^r_{2}]^{-1}$ is the
retarded Green's function of the central disordered SWNT and the
relationship between self-energy and linewidth function is
$\Gamma_{1,2}=i[\Sigma_{1,2}^{r} -\Sigma_{1,2}^{a}]$. With the help
of density of states given by Green's functions, we can express
emittance as \cite{grame97}
\begin{equation}
E_\mu=-\rm{Tr}( \rm{Im}[G^a\Gamma_{2}G^r\Gamma_{1}G^a]_{xx}
-\frac{D_{1xx}D_{2xx}}{D_{1xx}+D_{2xx}})
\label{e1}
\end{equation}
where $D_\alpha=G^r\Gamma_\alpha G^a$, $D_{\alpha
xx}=[D_\alpha]_{xx}$, and $\rm{[\cdots]_{xx}}$ denotes the diagonal
element of the relevant square matrix. The quantities in the first
and second terms in Eq.(\ref{e1}) have clear physical meaning. The
first term is the partial DOS $dN_{12}/dE$ (up to a sign) describing
the DOS of the electron coming from the second lead and exiting the
first lead while the second term consists of local
DOS\cite{buttiker4} and is due to the displacement current. The
dynamic response given by $E_\mu$ is either positive for small
transmission coefficient or negative for large transmission
coefficient giving rise to an inductive-like or capacitive-like
behavior. For instance, for small transmission coefficent
$dN_{12}/dE$ is very small and the sign of the emittance is
dominated by the second term which is positive.

\begin{figure}
\includegraphics[width=9cm]{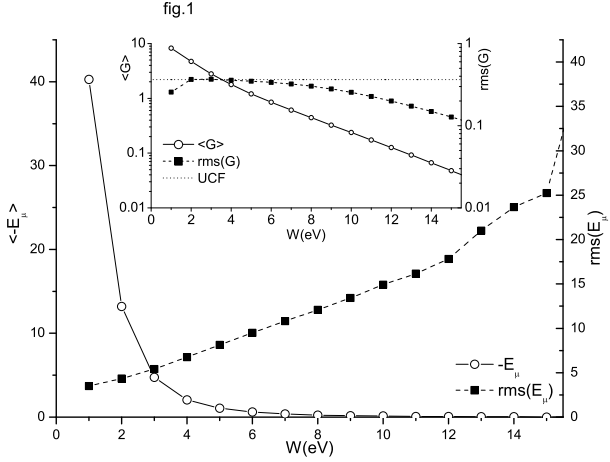}
\caption{The mean and rms values of emittance $-E_\mu$ at the
subband center E=3.2eV for different disorder strengths (one million
configurations averaged). Inset shows the corresponding mean and rms
values of conductance in unit $2e^2/h$.\label{fig4}}
\end{figure}

We first examine the emittance fluctuation defined by the root mean
square (rms) values as rms$(E_{\mu})=[\langle E_{\mu}^{2}
\rangle-\langle E_{\mu} \rangle ^{2}]^{1/2}$, where $\left\langle
{\cdots}\right\rangle $ denotes averaging over an ensemble of
samples, with different configurations of the same disorder
strength. Fig.1 shows the averaged emittance and its fluctuation vs
disorder strength with the energy fixed at the subband center. We
also plot the average conductance (with unit $2e^2/h$) and its
fluctuation for comparison.
As the disorder increases the system crosses over from
quasi-ballistic to diffusive regime.
Eventually for strong disorders the nanotube becomes an insulator in
the localized regime. The localization length $\xi$ defined as
$G=G_N \exp(-2L/\xi)$ is shown in Fig.4a where $G_N$ is the
conductance in the absence of disorder and $L$ is the system size.
We see that the conductance fluctuation in the diffusive regime
$W=[2,4]$ (where $\xi>L$) approaches to the one dimensional value of
UCF $0.73 e^2/h$ (see the dotted line in the inset of Fig.1)
obtained from the diagrammatic perturbation theory\cite{lee85} and
the random matrix theory\cite{beenakker}.

The main panel of Fig.1 shows that the average emittance is always
negative.
This is a very counter-intuitive result since we expect that a
conductor with small average conductance gives a capacitive-like
behavior corresponding to a positive emittance. It is true that for
any {\it specific configuration}, the conductance and emittance are
well correlated from our numerical result, i.e., large conductance
corresponds to negative emittance and small conductance corresponds
to positive emittance. The reason that the {\it average} emittance
gives an inductive-like response is due to the so-called necklace
states (or precursor of necklace states in the diffusive regime)
that are rare events in the disordered systems. Through these
states, electrons can tunnel through the disordered system of size
$L$ via multiple scattering. The larger the system size, the smaller
the number of necklace states is. The existence of these extended
states in localized regime was predicted by Pendry\cite{pendry}.
Different from the single resonant tunneling, the necklace state is
a necklace of n quasi-extended states stretching from one site of
the disordered sample to other. They occupy only a fraction of sites
with fractal dimension $1/2$ in 1D reminiscent of percolation
path\cite{pendry}. The analogous optical necklace states have been
observed experimentally in disordered multilayer
systems\cite{bertolotti} where the high transmission peaks show up
in the transmission spectrum while the average of logarithmic of
transmission coefficient decays linearly with the thickness.
Since the necklace states
(multi-resonant states) are rare events, they have very long life
time giving rise to a large negative emittance. To make this point
clear, we plot in Fig.2 the distribution functions $P(E_{\mu})$
versus emittance $E_{\mu}$ at different disorder strengths $W$. For
this figure we have used data from 200,000 different random
configurations except in panel (h). We see that at low disorder
strengths when $W\le 2$ the distribution is approximately Gaussian.
In this disorder range, as the disorder strength increases the
values of emittance are all negative and the mean emittance
gradually increases towards positive value. For larger disorder
strength $W=4$ in panel (d), the distribution is still approximately
Gaussian but the emittance can be positive while the mean emittance
is negative. When $W=5,6$ the distribution deviates gradually from
the Gaussian distribution and the mean emittance is around zero but
negative. At even stronger disorder $W=10$, the distribution is
sharply peaked at positive emittance with a long tail in the
negative emittance (see Fig.2h). To be more precise, the
distribution is asymmetric with large probability at small positive
emittance while the probability of negative emittance is small but
remains finite even for large negative emittance resulting a {\it
negative} averaged emittance.

\begin{figure}
\includegraphics[width=9cm]{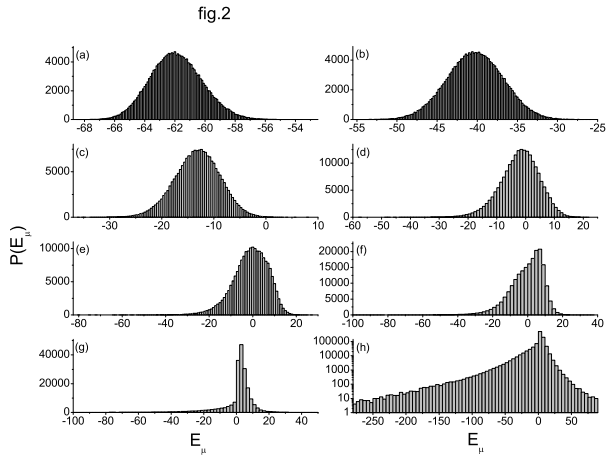}
\caption{The histograms of emittance at the subband center E=3.2 eV
for different disorder strength W=0.5, 1.0, 2.0, 4.0, 5.0, 6.0
and 10.0 eV, respectively. Panel (h) is the same as panel (g) but in
a log scale of $P(E_\mu)$ with one million configurations.\label{fig2}}
\end{figure}

Why the effect of necklace states has not been seen in the
conductance fluctuation but important for the dynamic response? This
is because at large disorders, there exists small number of necklace
states with small probability. For these necklace states the
magnitude of emittance can be very large while the largest possible
value of conductance is around one since it is unlikely for all
multichannels to tunnel through. Therefore the contribution of
necklace states to the average conductance is very small but is very
large for the average emittance. For instance, in Fig.2h with
$W=10$, the average emittance is $-0.11$ and total conductance from
eleven channels is $G=0.24$. Out of one million configurations, the
total probability of these necklace states with emittance
$E_\mu<-100$ is $0.3$ percent. Their contribution to the average
emittance and average conductance are, respectively, $E_\mu=-0.39$
and $G=0.002$. Clearly, these necklace states are important for
average emittance and negligible for average conductance. Without
these necklace states the average emittance would be positive. For
exactly the same reason, the emittance fluctuation is greatly
enhanced due to the existence of these necklace states. We point out
that these necklace states exist as long as the system size $L$ is
finite. In addition to the quasi-1D system, we have also studied the
averaged emittance of a disordered 1D tight-binding chain. Our
result indicates that our conclusion remains, i.e., the average
emittance is always negative for all disordered strengths.

Note that the necklace states exist in the localized regime and our
analysis above on these states is also done in the same regime. Now
we wish to argue that the average emittance $E_\mu$ is negative in
the diffusive regime as well. Since $E_\mu$ is a monotonic function
of disorder strength, hence $E_\mu$ in the localized regime is
larger than $E_\mu$ in the diffusive regime. The fact that $E_\mu$
is negative in the localized regime means that $E_\mu$ is negative
in all regimes including diffusive regime. Physically, in the
diffusive regime, the system is roughly described by a diffusive
conductor in parallel with multi-resonant channels connecting the
leads (the precursor of necklace states). It is the appearance of
these non-diffusive elements in the diffusive system, the precursor
of necklace states that changes the sign of $E_\mu$ in the diffusive
regime from positive to negative.

Now we study the emittance using a tight-binding model for a quantum
dot and a 2D system with disorders.
The system size is $L=40a$ where $a$ is the
lattice spacing and two electrodes of width $L_1$ are attached to
the scattering region. For the quantum dot $L_1=L/4$ while for the
2D system $L_1=L$. The average conductance and average emittance and
their fluctuations for the quantum dot and 2D system are depicted in
Fig.3 where we have collected one million configurations for each
data point. From Fig.3 we see that conductance fluctuation in the
diffusive regime exhibits plateaus with universal values that are
close to the theoretical predictions $UCF=0.70 e^2/h$ for quantum
dot\cite{beenakker} and $UCF=0.86 e^2/h$ for 2D system\cite{lee85}.
For emittance, our results show that for both the quantum dot and
the 2D system, the emittances are negative independent of disorder
strengths that disagrees with the theoretical predictions. The
emittance fluctuations for quantum dot (QD) and 2D system are
consistent with 1D case.

\begin{figure}
\includegraphics[width=9cm]{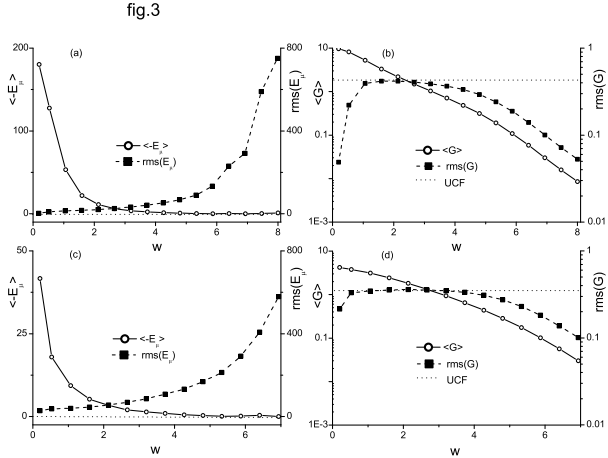}
\caption{The mean and rms values of emittance $-E_\mu$ of the 2D system in panel (a)
at $E=0.62$ and the QD in panel (c) at
$E=3.0$ for different $W$ (one million configurations
averaged). The corresponding mean and rms values of conductance in unit $2e^2/h$
are shown in panel (b)
and (d), respectively, for the 2D system and the QD.\label{fig3}}
\end{figure}

Our results for the average emittance clearly disagree with the
theoretical results obtained from the random matrix theory. This is
because in order to use analytic approach, one has to make the
following approximation on the second term in Eq.(\ref{e1}) so that
all the quantities involved are global,
\begin{equation}
\rm{Tr}[\frac{D_{1xx}D_{2xx}}{D_{1xx}+D_{2xx}}] \approx \frac{\rm{Tr}(D_1)\rm{Tr}(D_2)}
{\rm{Tr}(D_1+D_2)}
\label{appr}
\end{equation}
It is easy to see that this approximation greatly overestimates the
second term in Eq.(\ref{e1}) at large disorders. For weak disorders
with large transmission coefficient $D_{1xx}$ and $D_{2xx}$ are in
the same order of magnitude hence Eq.(\ref{appr}) is a good
approximation. However, at strong disorders there is a large
mismatch between $D_{1xx}$ and $D_{2xx}$ for quite a number of sites
$x$. Since $D_{1xx}D_{2xx}/(D_{1xx}+D_{2xx})\approx D_{2xx}$ when
$D_{1xx}>>D_{2xx}$ the left hand side of Eq.(\ref{appr}) is
approximately given by $\rm{Tr}[min(D_{1xx},D_{2xx})]$ which is much
smaller than the right hand side of Eq.(\ref{appr}) at strong
disorders. Using this approximation, our numerical results are shown
in Fig.4. We see that, indeed, for both quasi-1D and 2D systems the
average emittance with the approximation Eq.(\ref{appr}) (solid
square) overestimated the second term in Eq.(1) so that the results
are very different from that obtained using Eq.(1). This is because
the necklace states have been washed out in Eq.(\ref{appr}). In
addition, Fig.4 shows that using approximation Eq.(\ref{appr}) the
emittance is positive in the diffusive regime in agreement with the
theoretical predictions.

\begin{figure}
\includegraphics[width=8cm]{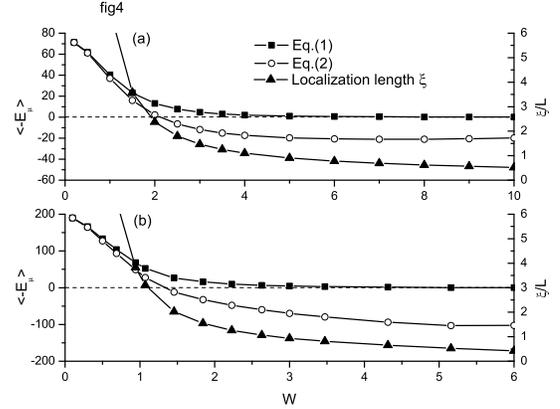}
\caption{The average emittance $-E_\mu$ and the localization length
$\xi/L$ vs $W$ ($100~000$ configurations averaged) for quasi-1D and
2D systems are shown, respectively, in panel (a) and (b). The other
parameters are the same as in Fig.1 and Fig.3a.\label{fig4}}
\end{figure}

In conclusion, we have studied the fluctuation of emittance and its
distribution for disordered 1D, 2D, and quantum dot systems.
Our numerical results indicate that the average emittance is always
negative independent of disorder strengths. Our findings disagree
qualitatively with the theoretical results obtained from  the random
matrix theory. The disagreement is due to the existence of
non-diffusive elements, necklace states or the precursor of necklace
states that are important for the dynamic response and negligible
for the conductance fluctuation.

\acknowledgments This work was financially supported by RGC grant
(HKU 7044/04P) from the government SAR of Hong Kong and LuXin Energy
Group. Endeavor Australia Cheung Kong Programme sponsored by Cheung
Kong Group is also acknowledged (W.R.).

$^*$ Present address: Department of Physics, The Hong Kong
University of Science and Technology, Clear Water Bay, HK, China

$^\dagger$ Electronic address: jianwang@hkusua.hku.hk

\end{document}